\documentclass[aps,prapplied,reprint,superscriptaddress,showpacs,showkeys]{revtex4-1}

\usepackage[multidot]{grffile}
\usepackage{amsmath,amssymb}  
\usepackage{color}
\usepackage{graphicx}

\usepackage{titlesec}
\titleformat*{\section}{\normalsize\bfseries\filcenter}
\titleformat*{\subsection}{\normalsize\bfseries\filcenter}
\titleformat*{\subsubsection}{\normalsize\itshape\filcenter}

\draft
\begin{document}

\title{How carrier memory enters the Haus master equation of mode-locking}

\author{Jan Hausen}
\email[]{hausen@campus.tu-berlin.de}
%\homepage[]{Your web page}
%\thanks{}

\affiliation{ Institute of Theoretical Physics, Technische Universit{\"a}t Berlin, Hardenbergstraße 36, 10623 Berlin, Germany}

\author{Kathy L{\"u}dge}
\email[]{kathy.luedge@tu-berlin.de}
\affiliation{ Institute of Theoretical Physics, Technische Universit{\"a}t Berlin, Hardenbergstraße 36, 10623 Berlin, Germany}

\author{Svetlana V. Gurevich}
\affiliation{Departament de F\'{\i}sica, Universitat de les Illes Balears \&
	Institute of Applied Computing and Community Code (IAC-3), Cra.\,\,de
	Valldemossa, km 7.5, E-07122 Palma de Mallorca, Spain}
\affiliation{Institute for Theoretical Physics, University of M{\"u}nster, Wilhelm-Klemm-Str. 9,
	48149 M{\"u}nster, Germany}

\author{Julien Javaloyes}
\affiliation{Departament de F\'{\i}sica, Universitat de les Illes Balears \&
	Institute of Applied Computing and Community Code (IAC-3), Cra.\,\,de
	Valldemossa, km 7.5, E-07122 Palma de Mallorca, Spain}

\begin{abstract}
We present a generalization of the Haus master equation in which a dynamical boundary condition allows to describe complex pulse trains such as the Q-switched and harmonic transitions of passive mode-locking as well as the weak interactions between localized states. As an example, we investigate the influence of group velocity dispersion on the stability boundaries of the Q-switched regime. We compare our results with that of a time-delayed system.
\end{abstract}

\maketitle

\emph{Mode-locking} (ML) is a well-known method to generate ultra stable
picosecond pulse trains with high repetition rates. It is a fundamental
component of many modern technologies ranging from telecommunications
towards medicine or spectroscopy~\citep{KT-PR-06}. The ML states
consist in the self-ordering of many lasing modes, 
thereby leading to regimes in which the temporal features
of the electric field intensity and of the active medium usually differ by several
orders of magnitudes. These difficulties explain why ML remains a
subject of intense research~\citep{AJ-BOOK-17} but also set limits
over the practical complexity of the modeling approaches.
The Haus master equation (HME) is a widely used model for active and
passive mode-locking that can be derived from general
principles using, in particular, the assumptions of small gain, losses
as well as weak spectral filtering \citep{HAU00,HAU75}. The HME 
materializes as a partial differential equation (PDE) in which
one adds all the physical effects influencing the field temporal profile.

In this letter we propose a generalization of the HME that allows
preserving carrier memory from one round-trip towards the next.
We can therefore access in an unified framework the regimes involving 
fast variations of the gain over the round-trip timescale $ \tau_c $ as well 
as those involving a slow evolution over time scales longer than the gain recovery $ \tau_g $.
Including carrier memory comes at a marginal computational
cost and consists in providing a dynamical equation for the gain at
the beginning of the round-trip. The latter is allowed to evolve on
a slow time scale and plays the role of a dynamical boundary condition.
Taking advantage of our formulation, we investigate the 
self-pulsing/Q-switch instability of passive mode-locking 
(PML), the occurence of harmonic mode-locking (HML) transitions, as well 
as the interactions between localized states.

In spite of its respectable age, the HME still holds salient advantages
as compared to more modern approaches based upon time delayed systems
(TDS)~\citep{VLA04,VLA05,MB-JQE-05,SCH19b,SCH19e,SJG-PRAp-20}. 
For instance, the inclusion chromatic dispersion is straightforward 
in a PDE framework, while it proved challenging in TDSs \citep{PSH-PRL-17,SCH19b}.
The HME also owns its computational efficiency to the fact that one restricts
the analysis of the field evolution to the vicinity of the pulse. 
Yet, this feature was recently borrowed from the HME and
applied to TDSs leading to a hybrid approach bridging
part of the divide between TDSs and the HME \citep{SCH18f}. However, the undeniable
intuitive power of the HME can be appreciated for instance studying
compound cavities where parasitic optical feedback \citep{JNS-PRE-16}
has a clear and interesting interpretation as non-local perturbations
leading, e.g., to exotic solitonic molecules \citep{JAV17}. 

The HME was originally designed for gain media that are slowly evolving on
the time scale of the cavity round-trip ($\tau_{g}\gg\tau_{c}$);
in this situation the gain temporal profile remains quasi uniform
within the cavity. However, strong gain depletion can be rigorously
included in the HME, starting from the Maxwell-Bloch equation \citep{PER20}.
How memory effects mediated by the incomplete recovery of the gain
influence pulse dynamics in the intermediate cavity regimes $\tau_{g}\sim\tau_{c}$
remains an open question. It is of particular interest in semiconductor lasers
for which $\tau_{g}$ is in the nanosecond range. For instance,
carrier memory is essential for the proper reproduction of the Q-switch 
mode-locking (QSML) instability of PML. There, the pulse train
gets strongly modulated over a slow time scale of several hundreds
of round-trips. The harmonic transitions of PML, in which the laser transitions 
from emitting $N$ towards $N+1$ pulses per round-trip, also depends on the 
carrier recovery time.  Finally, gain dynamics is essential for properly 
reproducing the interactions between the localized structures 
(LSs) observed in VECSELs~\citep{MJB-PRL-14}. 

However, the TDS descriptions of laser mode-locking~\citep{MB-JQE-05,VLA05} remain an
essential tool allowing not only for the analysis of pulsed regimes but 
also to unveil their connections with all the other possible solutions such as the continuous waves regimes. 
They were successfully extended to photonic crystals \citep{HBM-OE-10} and to the 
consideration of optical feedback, coherent optical injection \citep{AHP-JOSAB-16}, 
nonlocal imaging conditions \citep{MJB-JSTQE-15}, sub-threshold localized structures
(LSs) \citep{MJB-PRL-14} and complex geometries \cite{HML-PRAp-19,AP-MDPI-19}; most of these situations involve strong gain depletions, 
also termed coherent effects \citep{PER20}, leading to pulse shape parity symmetry 
breaking and drifts in semiconductor devices, either in active \citep{PER20} or 
passive \citep{JAV16a} ML configurations. 
The purpose of our amended Haus master equation is not to replace TDSs in the description of ML.
Rather, it offers a different point of view and allows further closing the gap with TDSs.
More specifically, we shall compare our result with those of the TDS model of PML 
that considers unidirectional propagation in a ring laser \citep{VLA05}. The equations
for the field amplitude $A$, the gain $G$, and the absorption $Q$ in the TDS
read: 
\begin{align}
\dot{A} & =-\gamma A+\gamma\sqrt{\kappa}e^{\frac{1}{2}[(1-i\alpha)G-(1-i\beta)Q]}A(t-\tau_c),\label{DDE_A}\\
\dot{G} & =\gamma_{g}G_{0}-\gamma_{g}G-e^{-Q}(e^{G}-1)|{A}|^{2},\label{DDE_G}\\
\dot{Q} & =\gamma_{q}Q_{0}-\gamma_{q}Q-se^{-Q}(e^{Q}-1)|A|^{2}\,.\label{DDE_Q}
\end{align}
Here, $\gamma_{g,q}$ are the carrier relaxation rates in the gain and absorber sections, $\gamma$ is the gain bandwidth, $G_0$, $Q_0$ are the unsaturated gain and absorption, $k$ is the cavity loss, $s$ is proportional to the ratio of the differential gain coefficients in the absorber and the gain, $\alpha$,$\beta$ represent the amplitude phase coupling in the active sections and $\tau_c = L/c$ is the cold cavity round-trip time.

\begin{figure}[t]
	\includegraphics{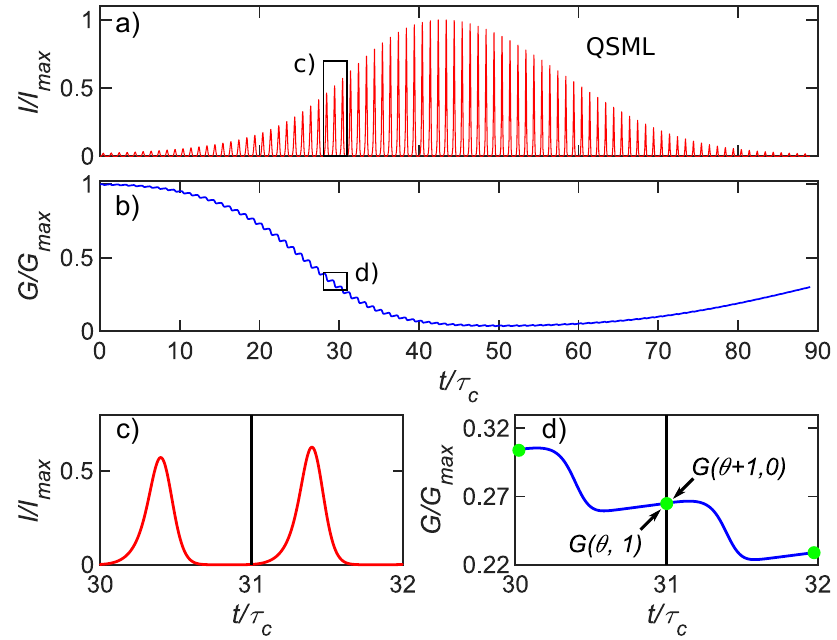}
	\caption{Time-series of the field intensity $I = |A|^2$ (a) and gain dynamics (b) in the quasi-periodic QSML regime over 90 round-trips. Black squares mark the close ups shown in panels c) and d), which correspond to two cavity round-trips with the border between two integration domains indicated by the black vertical line. Other parameters ($\gamma_g, \tau_c, \gamma, Q_0, s, \alpha, \beta) = (0.01, 2,40,0.3, 30,1.5,1.0)$ are normalized to $\gamma_q$.} 
	\label{Fig1}
\end{figure}

%%%%%%%%%%%%%%%%%%%%%%%%%%%%%%%%%%%%%%%%%%%%%%%%%%%%%%%%%%%%%%%%%%%%%%%%%%%%%%%%%%%%%%%%%%%%%%%%%%%%%%%%%%%%%%%%%%%%%%%%%%%%%%%%%%%%%%%%%%%%%%%%%%%%%%%%%%%%%%%%%%%%%%%%%%%%%%%%%%%%%%%%%%%%%%%%%%%%%%%%%%%%%%%%%%%%%%%%%%%%%%%%%%%%%%%%%% Model part
There are different dynamical regimes in which long-term carrier dynamics are crucial to correctly model the pulse dynamics. An example of such a regime is QSML, see Fig.\ref{Fig1}(a,b). As visible from the close ups in Fig.\ref{Fig1}(c,d), two consecutive pulses do not equally deplete the gain and therefore the gain value at the beginning of the consecutive domains (marked by green circles in Fig.\ref{Fig1}(d)) varies but is periodic on a time-scale of $\approx 160$ round-trips. From the characteristics of the gain dynamics in Fig.\ref{Fig1}(d), one intuitively recognizes that an asynchronous boundary condition (BC) simply represents the continuity of the solution \cite{GIA96}
\begin{align}
G(\theta, 1) &= G(\theta+1, 0), \label{async}
\end{align}
with $\theta$ being an integer describing the evolution from one round-trip towards the next; it represents a slow time-scale while the second variable $\sigma$ refers to the evolution over the fast time-scale within one round-trip.
Our objective is to formulate a model that includes these long term carrier correlations without the inconveniences of an asynchronous BC. Notice that in \citep{PER20} the authors solved this problem by adding a linear integro-differential equation describing the gain modulation on the fast time scale to complement the standard HME. Here, instead, we propose to transform the continuity equation Eq.~(\ref{async}) into a dynamical boundary condition. 
We start by deriving the standard HME from the TDS~(\ref{DDE_A}-\ref{DDE_Q}). This can be achieved using a two time-scale approach in the uniform field limit~\cite{HAU00,KOL06,CAM16} leading to:
\begin{align}
\label{HAUS_A}
\partial_{\theta} A &=\frac{1}{2(\tau_c\gamma)^2} \partial_{\sigma}^2 A +\frac{1}{2} [(1-i\alpha)G-(1-i\beta)Q-k]A,\\
\label{HAUS_G}
\partial_{\sigma} G &=  \gamma_g G_0-\gamma_g G-G|{A}|^2,\\ 
\label{HAUS_Q}
\partial_{\sigma} Q &=  \gamma_q Q_0-\gamma_qQ-s Q|A|^2.
\end{align}
Equations (\ref{HAUS_A})-(\ref{HAUS_Q}) describe the electric field evolution from round-trip to round-trip $\partial_{\theta} A$, and the variation of the carrier densities $G$ and $Q$ on the fast time-scale $\sigma$. To maintain the long-term carrier dynamics we reformulate the BC Eq.~(\ref{async}) as an \emph{a synchronous} ordinary differential equation; we first integrate the gain Eq.~(\ref{HAUS_G}) over the cavity length in $\sigma$ and then insert Eq.~(\ref{async}) in the left-hand side, yielding:
\begin{equation}
G(\theta+1,0)-G(\theta,0) = \gamma_g G_0- \int_{0}^{1}\gamma_g G-G|{A}|^2 d\sigma.
\end{equation}
We define the gain value at the beginning of each round-trip as $G(\theta,0) = \mathcal{G}(\theta)$ and assume a slow evolution between consecutive round-trips such that $\mathcal{G}(\theta+1)-\mathcal{G}(\theta) \approx d\mathcal{G}/d\theta$ (note that higher order approximations are possible). This leads to the following dynamical equation for the boundary condition complementing the HME system (\ref{HAUS_A}-\ref{HAUS_Q}):
\begin{equation}
\label{finBC}
\frac{d \mathcal{G} }{d\theta} =  \gamma_g G_0- \int_{0}^{1}\gamma_g G-G|{A}|^2 d\sigma.
\end{equation}
An identical treatment is performed for $Q(\theta,0) = \mathcal{Q}(\theta)$ and we 
integrate the generalized HME system~(\ref{HAUS_A}-\ref{HAUS_Q},\,\ref{finBC})
using $(\mathcal{G},\mathcal{Q})$ as inhomogeneous Dirichlet boundary conditions 
on the left side of the cavity. The resulting PDEs were solved using standard split-step 
methods such as outlined in~\cite{GUR17}. As previously mentioned, strong gain depletion 
induces parity symmetry breaking in the pulse shape and a slow drift of the pulse within 
the cavity. Adding a translation operator $(\upsilon \partial_\sigma)$ in (\ref{HAUS_A}) 
conveniently freezes the pulses position in the cavity. The value of $\upsilon$ is 
simply evaluated by calculating the residual motion after each round-trip.
\begin{figure}[t]
	\includegraphics{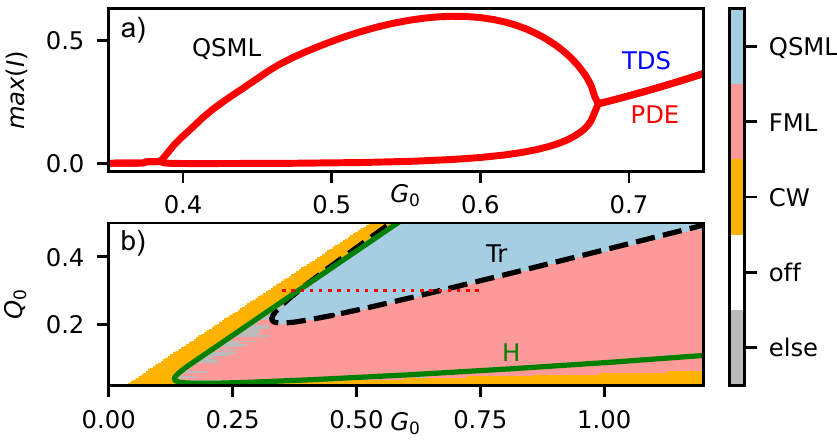}
	\caption{(a) 1D-bifurcation diagram with the maximum intensity $\max(I)$ as a function of the pump $G_0$ showing the QSML regime (cf. Fig.\ref{Fig1}). For the TDS (\ref{DDE_A}-\ref{DDE_Q}) (blue) all pulse intensities obtained in 500 round-trips are plotted, for the PDE (\ref{HAUS_A}-\ref{HAUS_Q},\,\ref{finBC}) only the maximum and minimum pulse intensity are indicated by red lines. (b) 2D-bifurcation diagram showing the different dynamic regimes in the ($G_0$,\,$Q_0$) plane obtained integrating the HME~(\ref{HAUS_A}-\ref{HAUS_Q},\,\ref{finBC}) with fundamental mode-locking (FML), Q-switched mode-locking (QSML) and continuous wave operation (CW). The black dashed line marks a Torus (Tr) and the green line a Hopf (H) bifurcations, both obtained from the TDS. The dotted red line marks the $Q_0$ value used in (a). Parameters are as in Fig.~\ref{Fig1}.} 
	\label{Fig2}
\end{figure}
%%%%%%%%%%%%%%%%%%%%%%%%%%%%%%%%%%%%%%%%%%%%%%%%%%%%%%%%%%%%%%%%%%%%%%%%%%%%%%%%%%%%%%%%%%%%%%%%%%%%%%%%%%%%%%%%%%%%%%%%%%%%%%%%%%%%%%%%%%%%%%%%%%%%%%%%%%%%%%%%%%%%%%%%%%%%%%%%%%%%%%%%%%%%%%%%%%%%%%%%%%%%%%%%%%%%%%%%%%%%%%%%%%%%%%%%%% Model verifcation part

Utilizing the generalized HMEs.~(\ref{HAUS_A}-\ref{HAUS_Q},\,\ref{finBC}) we can reconstruct the QSML instability as shown in Fig.~\ref{Fig2}(a), in which the maximum and minimum pulse intensity found in 500 round-trips for different values of the pump $G_0$ are indicated in red. 
We use a standard set of parameters similar to \cite{SCH18f}, and given in the caption of Fig.~\ref{Fig1}. 
For comparison, all intensity maxima obtained from TDS~(\ref{DDE_A}-\ref{DDE_Q}) are presented in blue. One can clearly see that the onset and disappearance of the QSML is well preserved by the generalized HME~(\ref{HAUS_A}-\ref{HAUS_Q},\,\ref{finBC}), since the correlations between the gain and the field intensity from one round trip to the next are properly accounted for. Furthermore, we investigate the bifurcation boundaries of the QSML regime. Therefore, we numerically integrate the PDE system in the ($G_0,\,Q_0$)-plane sweeping along $G_0$ and additionally acquire the two-parameter bifurcation lines representing the domain boundaries obtained from the TDS utilizing the path continuation software ddebiftool \cite{DDEBT}. We find that the torus bifurcation (Tr) bounding the QSML regime \cite{VLA05}, indicated by the black dashed line, matches the numerical result generated using the HME model (light blue region). Moreover, the Andronov-Hopf bifurcation (green line H) representing the onset of the fundamental mode-locking regime (FML) is correctly described by the HME with the dynamical boundary condition.

\begin{figure}[t]
	\includegraphics{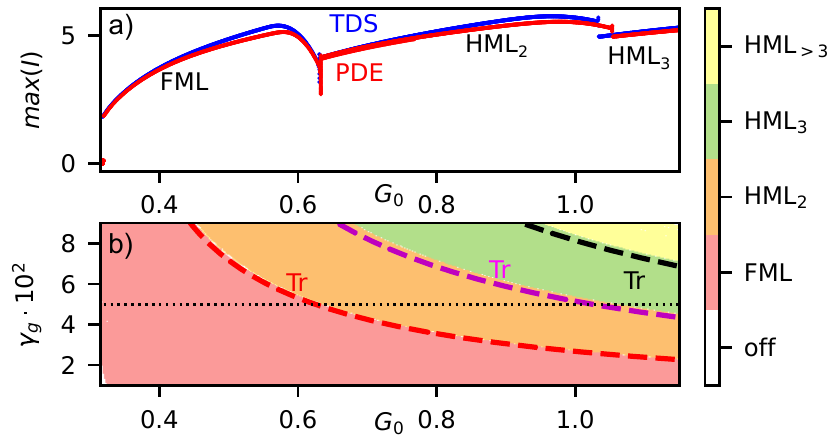}
	\caption{(a) 1D-bifurcation diagram in $G_0$ across the FML regime, for the TDS~(\ref{DDE_A}-\ref{DDE_Q}) (blue) and HME~(\ref{HAUS_A}-\ref{HAUS_Q},\,\ref{finBC}) (red) all pulse intensities obtained in 250 round-trips are plotted. (b) 2D-bifurcation diagram of the dynamics in the ($G_0$,\,$\gamma_g$) plane obtained by integrating the HME. The colors indicate the dynamical regimes with HML$_n$ referring to harmonic mode-locking with $n$ pulses. The dashed lines represent the Torus bifurcation lines (Tr) bounding the FML (red), HML$_2$ (magenta) and HML$_3$ (black) regimes in the TDS, obtained using ddebiftool. The dotted black line marks the $\gamma_g$ value used in (a). Parameters are: ($\tau_c, \gamma, Q_0, s, \alpha, \beta) = (30,10,0.3, 3,0,0)$.} 
	\label{Fig3}
\end{figure}

Similarly, we show that the transition between FML and harmonic mode-locking (HML) can also be modeled utilizing the generalized HME as indicated in Fig.~\ref{Fig3}. This transition is characterized by the emergence of an additional pulse, which becomes stable at an equilibrium distance such that the pulses deplete the gain equidistantly and are hence of the same amplitude.
In Fig.\ref{Fig3}(a) we compare the unique intensity maxima found in 250 round-trips in the TDS (blue) and HME (red) and in (b) a two-parameter bifurcation diagram of the dynamics in the ($G_0$,\,$\gamma_g$) plane is shown, together with bifurcation lines obtained using ddebiftool. The upper stability boundaries of the FML and HML$_2$ regime are given by torus bifurcations along the respective solution branches and are indicated by dashed lines. They match very well the boundaries obtained by integrating the generalized HME.

\begin{figure}[t]
	\includegraphics{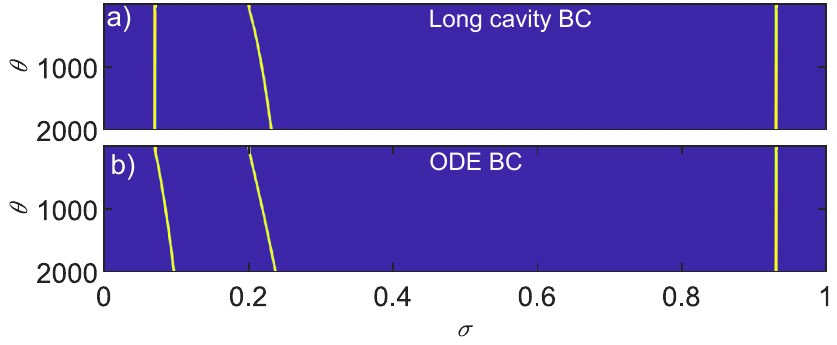}
	\caption{Two time-scale representation of the transient dynamics of the intensity in the localized state regime, with the fast time-scale on the x axis $\sigma$ and the slow time-scale on the y-axis ($\theta$) describing the number of round-trips. a) Long cavity boundary condition with $\mathcal{G}(\theta) = G_0$, b) dynamical BC as in Eq.~(\ref{finBC}). Parameters as in Fig.~\ref{Fig3}, with $\tau_c = 400$ and $\gamma_G = 0.04$.} 
	\label{Fig4}
\end{figure}

%%%%%%%%%%%%%%%%%%%%%%%%%%%%%%%%%%%%%%%%%%%%%%%%%%%%%%%%%%%%%%%%%%%%%%%%%%%%%%%%%%%%%%%%%%%%%%%%%%%%%%%%%%%%%%%%%%%%%%%%%%%%%%%%%%%%%%%%%%%%%%%%%%%%%%%%%%%%%%%%%%%%%%%%%%%%%%%%%%%%%%%%%%%%%%%%%%%%%%%%%%%%%%%%%%%%%%%%%%%%%%%%%%%%%%%%%% LS Transients
Another effect which requires to model the evolution of the gain at the beginning of each round-trip is the interaction of pulses via the tails of the exponential gain relaxation, if they are not in an equilibrium state (i.e. do not deplete the gain equidistantly). This becomes important if several pulses are excited in a long external cavity ($\tau_c \gg \tau_g$) mode-locked laser, so that they become temporally localized \cite{MAR14c}. This scenario is depicted in Fig.\ref{Fig4}, where we show the transient behavior of the intensity of three pulses in the pseudo space-time representation~\cite{GIA96}. The difference in the transients in Figs.~\ref{Fig4}(a) and (b) lies within the applied BCs. In Fig.~\ref{Fig4}(a), the long-cavity BC $\mathcal{G}(\theta) = G_0$ is used whereas for Fig.~\ref{Fig4}(b) the dynamical BC given by Eq.~(\ref{finBC}) is applied. The behavior of the first pulse, which does not change its position in (a), indicates that by just implementing the long cavity BC, the repelling interaction of the third pulse is lost. However, utilizing the BC Eq.~(\ref{finBC}) correctly models the gain relaxation at the domain boundary and therefore the drift of the first pulse is recovered.
%
%%%%%%%%%%%%%%%%%%%%%%%%%%%%%%%%%%%%%%%%%%%%%%%%%%%%%%%%%%%%%%%%%%%%%%%%%%%%%%%%%%%%%%%%%%%%%%%%%%%%%%%%%%%%%%%%%%%%%%%%%%%%%%%%%%%%%%%%%%%%%%%%%%%%%%%%%%%%%%%%%%%%%%%%%%%%%%%%%%%%%%%%%%%%%%%%%%%%%%%%%%%%%%%%%%%%%%%%%%%%%%%%%%%%%%%%%% GVD

\begin{figure}[b]
	\includegraphics{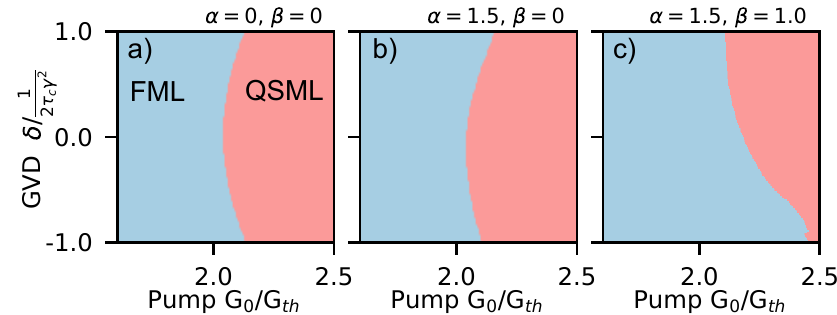}
	\caption{Dynamic regimes as a function of the group velocity dispersion (GVD) and pump $G_0$ (downsweep in $G_0$) in the QSML (red) regime at the transition to FML (blue) dynamics. a) b) and c) the results for different values of $\alpha,\beta$. Other parameters as in Fig. \ref{Fig1}.} 
	\label{Fig5}
\end{figure}

A further advantage of the HME is the comparably simple implementation of dispersive effects, which might be demanding utilizing TDSs~\cite{PIM17,SCH19b}. The influence of dispersion can be important when optimizing the laser performance~\cite{WAL16}, investigating instabilities \cite{PIM17,SCH19b}, or soliton formation. On the account of studying the effect of the group velocity dispersion (GVD) on the PML regime, we add an imaginary contribution  $(i\delta\partial_{\sigma}^2)$ to the electric field equation (\ref{HAUS_A}). Hence, we are able to investigate the influence of the GVD on the QSML boundaries, which has not been shown so far to the best of our knowledge. The interplay between the nonlinear influence of the $\alpha$ factors and the GVD is indicated in Fig.~\ref{Fig5}. Here down-sweeps along the pump $G_0$ for different values of the GVD were calculated and the resulting dynamics is presented in the color-code as in Fig.~\ref{Fig2}. In case of $\alpha = \beta = 0$ shown in Fig.\ref{Fig5}(a), the boundary between FML and QSML regimes shifts symmetrically around $\delta = 0$ to higher $G_0$; for large absolute values of the GVD the pulse is chirped and its excess bandwidth incurs more spectral filtering. 
The energy drops and therefore the gain is depleted less efficiently, making the FML only stable at higher pump powers. The symmetry in $\delta$ can be explained by the complex conjugate symmetry of eq.(\ref{HAUS_A}) when $\alpha = \beta = 0$. However, this changes if $\alpha \neq 0$ as depicted in Fig.~\ref{Fig5}(b). Here the interplay between the GVD and the nonlinear influence induced by $\alpha$ leads to a slight asymmetry in the boundary. This effect is strongly enhanced if also $\beta \neq 0$ as presented in Fig.~\ref{Fig5}(c). Due to the combined nonlinear effects, the pulse energy in the FML regime for positive GVD is higher at lower pump values $G_0$ as compared to negative GVD. Therefore, the gain depletion is larger/energetically more favorable, which makes the FML unstable at lower pump powers as compared to negative GVD.
%
%%%%%%%%%%%%%%%%%%%%%%%%%%%%%%%%%%%%%%%%%%%%%%%%%%%%%%%%%%%%%%%%%%%%%%%%%%%%%%%%%%%%%%%%%%%%%%%%%%%%%%%%%%%%%%%%%%%%%%%%%%%%%%%%%%%%%%%%%%%%%%%%%%%%%%%%%%%%%%%%%%%%%%%%%%%%%%%%%%%%%%%%%%%%%%%%%%%%%%%%%%%%%%%%%%%%%%%%%%%%%%%%%%%%%%%%%% Coonclousion
\ \\ In conclusion we have developed a generalization of the Haus master equation that includes a 
dynamical boundary condition, which makes it possible to study non-periodic regimes such as QSML, 
harmonic transitions, or transients of localized states. Especially in the intermediate cavity regime, 
this was not possible before with the standard HME model. Taking advantage of the PDE formulation, we are 
able to discuss the influence of the group velocity dispersion on the QSML instability threshold. 
Importantly, the presented model is potentially compatible with the path-continuation 
methods adapted to PDEs such as \cite{pde2path} and, as compared with TDS approaches, 
the generalized HME allows to take one step further in the analysis of ML. 
It is because the regularly pulsating ML solutions are periodic solutions of TDSs while 
they are steady states of the HME. Hence, one could access the quasi-periodic 
regimes of ML, such as e.g. breathing and modulated regimes, as periodic orbits 
of the HME; while the bifurcation analysis of periodic solutions of PDEs is 
feasible \cite{pde2path}, the analysis of quasi-periodic regimes 
in TDS is beyond the reach of current frameworks \cite{DDEBT}. Finally, including 
transverse diffraction in the generalized HME would allow the study spatio-temporal 
instabilities such as pulse filamentation of light bullets. 
\clearpage
\section*{Funding}
J.H., and K. L. thank the Deutsche Forschungsgemeinschaft (DFG) within the frame of the SFB787 and the SFB910 for funding. J.J. acknowledge the financial support of the MINECO Project MOVELIGHT (PGC2018-099637-B-100 AEI/FEDER UE). S.G. acknowledges the PRIME program of the German Academic Exchange Service (DAAD) with funds from the German Federal Ministry of Education and Research (BMBF).

\end{document}